\documentstyle[twocolumn,floats,epsfig,aps,prb]{revtex}

\begin{document}
\widetext
\draft
\twocolumn[\hsize\textwidth\columnwidth\hsize\csname
@twocolumnfalse\endcsname
\title{Jordan-Wigner approach to dynamic correlations in spin-ladders}
\author{Tamara S. Nunner and Thilo Kopp}
\address{EP VI, Universit\"{a}t Augsburg, 86135 Augsburg, Germany}
\date{\today}
\maketitle
\begin{abstract}

We present a method for studying the excitations of low-dimensional
quantum spin systems based on the Jordan-Wigner transformation. Using
an extended RPA-scheme we calculate the correlation
function of neighboring spin flips which well approximates the optical 
conductivity of ${\rm Sr_2CuO_3}$. We extend
this approach to the two-leg $S=\frac{1}{2}$--ladder by numbering the
spin operators in a meander-like sequence. We obtain good agreement
with the optical conductivity of the spin ladder compound
(La,Ca)$_{14}$Cu$_{24}$O$_{41}$ for polarization along
the rungs. For polarization along the legs higher order correlations
are important to explain the weight of high-energy
continuum excitations and we estimate the contribution of 4-- and
6--fermion processes. 
\end{abstract}
]
\narrowtext

\section{Introduction}
Due to the presence of strong quantum fluctuations, low-dimensional
spin systems show very complex behavior and provide a challenge 
for theoretical treatments. In this context, S=$\frac{1}{2}$--Heisenberg
spin ladders are especially interesting because they represent an
interjacent system, in between  the antiferromagnetic
S=$\frac{1}{2}$--Heisenberg  chain and the two-dimensional antiferromagnetic
Heisenberg model. Early on, these spin ladders were considered as
systems which display a dimensional crossover between one and two
dimensions.\cite{DagottoRice}
However, spin ladders do not constitute a ``smooth crossover'', because
even-leg spin ladders have a spin liquid ground state and finite spin
gap.~\cite{Rice,Azuma}
This is in contrast to the critical systems, the S=$\frac{1}{2}$--chain 
and the odd-leg ladders, which have algebraically decaying spin 
correlations and to the 2D antiferromagnetic Heisenberg model, 
for which a long-range N\'eel-ordered ground state was established.~\cite{CHN}
In a field theoretical mapping of the low-energy modes on a O(3)
nonlinear $\sigma$ model this even-odd effect in the number of legs 
results from the addition of a topological term. The term was
demonstrated to be  zero for even-leg ladders and the two-dimensional 
Heisenberg model and finite for odd-leg ladders as well as the
S=$\frac{1}{2}$--chain.\cite{Khvesh,Sierra,Chakra,ChakraHalperinNelson}
Correspondingly the system is gapless for odd-leg ladders and gapful
for even-leg ladders whereby the gap decreases exponentially with the number of legs.

In this article we will focus on the antiferromagnetic two-leg
S=$\frac{1}{2}$-ladders.
The two-leg ladder can be approached conceptionally from the limit of
strong coupling $J_\perp$ along the
rungs.~\cite{SachdevBhatt,GopalanRice,SushkovKotov,monienPRL} 
Then the elementary excitations can be considered as excitations of
rung-triplets which propagate throughout the ladder due to the
finite leg coupling $J$. 
For small coupling $J_\perp/J$, a more natural description would seem
to be in terms of the spinon excitations of the isolated legs.
However, the excitations of the two-leg spin-ladders cannot be
constructed perturbatively from spinons of the chains since the
rung coupling is a relevant perturbation. The spinons are
confined and have to form bound states in the
ladder-case.~\cite{SheltonNersesyanTsvelik,Greiter}

Of particular interest is the intermediate
coupling regime $J \approx J_\perp$, as this
case is related to the two-dimensional systems. It is also realized in
cuprate spin ladder compounds,~\cite{DCJohnston00,PRL2001,condmat}
which are of interest due to their affinity to the cuprate high-$T_c$-superconductors.
This intermediate regime is difficult to obtain 
and requires tedious calculations.

Here we propose an alternative
approach, which is based on the Jordan-Wigner transformation~\cite{JW28,JWdetails} and which
is fairly simple.
In contrast to previous applications of the Jordan-Wigner
transformation
to spin ladders by  Dai and Su~\cite{DaiSu}  and Azzouz {\it et al.}~\cite{Azzouz} 
we do not treat the phase
factor within a mean-field or a flux-phase approximation.
Rather we expand the phase factor and treat the resulting interaction terms
on the same footing as the Ising term, which corresponds to a
4-fermion term in the Jordan-Wigner representation.
We also apply this approach to the calculation
of dynamic correlation functions. In particular we discuss the 
optical conductivity $\sigma(\omega)$, as it provides valuable information about
high-energy $S$=0--spin excitations. In $\sigma(\omega)$
magnetic excitations can be observed  due to the simultaneous excitation of a
phonon, which breaks the inversion symmetry
between two neighboring spins. The mechanism, which has been suggested
by Lorenzana and Sawatzky,~\cite{LorenzanaSawatzky} provides a finite dipole moment
to a simultaneous flip of neighboring spins.
Since the phonon can take arbitrary momentum in this process,
the magnetic $S$=0--excitations have to be averaged over
the whole Brillouin zone to ensure zero total momentum.
This mechanism has recently been used to investigate the high-energy
spin excitations in the spin-ladder compound
(La,Ca)$_{14}$Cu$_{24}$O$_{41}$ by Windt {\it et al.}~\cite{PRL2001}
Using a perturbative approach, based on a
continuous-unitary-transformation~\cite{PRLKoeln}
and a Jordan-Wigner representation it was possible to identify the 
$S$=0--bound state of two triplets in the optical conductivity of 
(La,Ca)$_{14}$Cu$_{24}$O$_{41}$.

Here, we discuss in more detail the Jordan-Wigner approach
which we have previously used in Ref.~\onlinecite{PRL2001}.
We introduce our approximation scheme for the Jordan-Wigner fermions
for the simpler case of a 1d spin-chain in Sec.~\ref{sec:1d}.  
We calculate $\sigma(\omega)$  for the 1d-spin chain
in RPA and compare with the optical conductivity  of
${\rm Sr_2CuO_3}$ measured by Suzuura {\it et al.}~\cite{Suzuura}
In Sec.~\ref{sec:JWladder} we
generalize our Jordan-Wigner treatment to two-leg ladders by 
arranging the spins in a meander-like sequence. The mean field
treatment of this representation, which is based on the expansion of all phase factors, is
discussed in Sec.~\ref{sec:MeanFieldLadder} and the RPA-treatment
in Sec.~\ref{sec:RPALadder}. We calculate dynamic spin-flip correlation
functions in Sec.~\ref{sec:SpinFlipCorrFuncLadder} and compare
$\sigma(\omega)$ for the two-leg ladder with the optical conductivity of 
(La,Ca)$_{14}$Cu$_{24}$O$_{41}$ in Sec.~\ref{sec:SigmaLadder}. 
The weight of processes involving the excitation of more than
two fermions is estimated in Sec.~\ref{sec:HigherOrderContributions}.
In Appendix~\ref{app:DimLimit} it is shown that our mean field
treatment reproduces the correct strong coupling limit, 
in Appendix~\ref{app:RPAequations} the full set of RPA-equations
for the two-leg ladder is listed.
In Appendix~\ref{app:PhaseFactor} the role of the phase factor is
explored, and in Appendix~\ref{app:SymmetryBreaking} the consequences 
of the artificial symmetry breaking introduced by our mean field
treatment are discussed.

\section{Jordan-Wigner transformation for the 1d spin chain}
\label{sec:1d}

First we recall the Jordan-Wigner transformation for the 1d spin chain
and introduce our approximation scheme. The
fact, that we find fairly good agreement with the spinon 
evaluation for the 1d spin-chain, inspired us to extend
the approach to two-leg spin-ladders in Section~\ref{sec:JWladder}.

As spin operators do not obey canonical commutation
relations, they are often transformed into either
bosonic or fermionic operators, in order to permit the application of
standard diagrammatic perturbation theory.  
With any  mapping, however, the algebra of the original spin operators 
has to be preserved. For the Jordan-Wigner 
transformation this is provided by 
rewriting the spin operators as fermionic operators with a
long-ranged phase factor
\begin{equation}
S_i^-=c_i e^{i \phi_i} \, , \,\, \phi_i= \pi \sum_{j<i} c_j^\dagger
c_j \, , \,\, S_i^z=(c_i^\dagger c_i - \textrm{\small $\frac{1}{2}$}) \,.
\end{equation}

The representation is particularly useful for models with 
nearest-neighbor exchange interactions
because in products of neighboring spins the phase factors drop out. 
This is due to the fact that the
fermion operator $c_i$ commutes with the phase $\phi_i$ of the same
site ($[c_i,\phi_i]=0$) .
Therefore the 1d antiferromagnetic Heisenberg Hamiltonian transforms to 
\begin{eqnarray}
\label{eq:Hamilton1D}
H &=& J \sum_i \left \{ \frac{1}{2}
\left (c_i^\dagger c_{i+1} + c_{i+1}^\dagger c_i \right) \right.
\nonumber \\
& & \qquad\quad  + \left. \left( c_i^\dagger c_i - \frac{1}{2}\right)
                 \left(c_{i+1}^\dagger c_{i+1} - \frac{1}{2} \right) \right \}\, .
\end{eqnarray}
The first term in Eq.~(\ref{eq:Hamilton1D}) corresponds to
the XY-part of the original Heisenberg Hamiltonian. In the fermionic
representation it acts as a nearest neighbor hopping. 
The second term, which originates from
the Ising term of the original Heisenberg Hamiltonian,
introduces a nearest neighbor density-density interaction among the fermions.

\subsection{Mean field treatment -- d=1}

\begin{figure}[b!]
\begin{center}
\epsfig{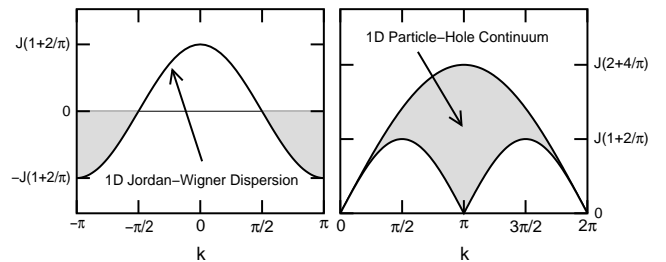}
\end{center}
\caption{Left panel: mean field dispersion for the Jordan-Wigner
fermions in $d$=1, grey shading denotes the filling
in the ground state. Right panel: continuum of Jordan-Wigner
particle-hole excitations in $d$=1.}
\label{fig:DispCont1D}
\end{figure}

Following Wang~\cite{YRWang92}, the Ising-interaction can be
treated in mean field approximation (MFA) by introducing
a nearest neighbor ``covalent bonding'' of the Jordan-Wigner
fermions $\chi=\langle c_i^\dagger c_{i+1} \rangle$.
\begin{equation}
\label{eq:1DMeanField}
H_{MF} = J \sum_k (1-2 \chi) \cos k \, c_k^\dagger c_k
\quad \textrm{with} \quad \chi=-\frac{1}{\pi}
\end{equation}
The ground state of the Heisenberg model has no net
magnetization $\langle S_i^z \rangle = 0$. Within the
fermionic representation this 
implies that the fermion system is at half filling
$\langle c_i^\dagger c_i \rangle = \frac{1}{2}$.
The ground state is obtained by filling up all negative energy
states. This leads to a Fermi surface at
wave vectors $k_f=\pm \frac{\pi}{2}$, as displayed in Fig.~\ref{fig:DispCont1D}.
Adding/removing a fermion to the system corresponds to 
$S_z=\pm 1$--excitations,
$S_z=0$--excitations can be realized by particle-hole excitations. 
The particle-hole continuum of the Jordan-Wigner fermions, displayed in
Fig.~\ref{fig:DispCont1D}, is very similar to the two-spinon
continuum~\cite{Yamada}. 
The upper cutoff of the Jordan-Wigner particle-hole continuum is at
$(2+\frac{4}{\pi})J \approx 3.27 J$ and therefore 
close to $\pi J$ which is the maximum energy for two spinons.

\subsection{RPA for optical conductivity -- d=1}

High-energy spin excitations can be observed in the mid-infrared range of
the optical conductivity $\sigma(\omega)$
due to the simultaneous excitation of a
phonon.~\cite{LorenzanaSawatzky} The optical conductivity of 
the 1d spin-chain compound ${\rm Sr_2CuO_3}$~\cite{Suzuura} has been
nearly perfectly reproduced by Lorenzana and Eder~\cite{LorenzanaEder}
using an ansatz based on numerical results in finite
chains, sum rules and  Bethe ansatz results.
Originally, a similar procedure was suggested
by M\"uller {\it et al.}\cite{Mueller} for the evaluation of the
dynamic structure factor $S(k,\omega)$, 
taking advantage of the observation that the  two-spinon contribution is 
the class of Bethe-ansatz solutions which carries most of
the weight of the continuum excitations. Only recently, it has been
possible to determine the two-spinon contribution to $S(k,\omega)$ 
exactly.\cite{Boug,Karbach}

For the optical conductivity, however, an exact expression of the two-spinon 
contribution is not yet available.
Nevertheless, the evaluation of Lorenzana and
Eder,~\cite{LorenzanaEder} which nearly perfectly reproduces
the shape of the cusp-like, wide structure in  $\sigma(\omega)$,
confirms that the observed resonance indeed results from
two-spinon excitations of the nearest-neighbor Heisenberg model.  
This motivated us to use the established $\sigma(\omega)$ of the 1d 
spin-chain as a reference and check for the quality of the results of
our analytical Jordan-Wigner approach. We calculate the two-particle 
correlation function $\sigma(\omega)$ within an extended 
RPA-scheme, i.e., by summing up bubble- and ladder-diagrams and compare
our result with the experimental optical conductivity of
${\rm Sr_2CuO_3}$.~\cite{Suzuura}

For the one-dimensional spin-chain the phonon-assisted magnetic
contribution to the optical conductivity is given
by~\cite{LorenzanaSawatzky,LorenzanaEder}
\begin{equation}
\sigma(\omega) \sim - 16 \,
\omega \sum_p \sin^4 (\frac{p}{2})
{\rm Im} \langle \langle \delta B_{-p}; \delta B_p
\rangle \rangle_{( \omega- \omega_{ph})} \, .
\label{eq:sigma1d}
\end{equation}
The spin-flip operator $\delta B_p$ is expressed in MFA by
\begin{eqnarray}
&\delta B_{p}& = \frac{1}{N}  \sum_i e^{ipr_i} \left (
       {\bf S}_{i} {\bf S}_{i+1}
        -  \langle {\bf S}_{i} {\bf S}_{i+1} \rangle
       \right ) \nonumber \\
& & \approx e^{-ip/2} \frac{1}{N} \sum_k (1-2 \chi)
 \cos (k+\frac{p}{2}) c_k^\dagger c_{k+p} \,.
\end{eqnarray}
This yields for the 
dynamic spin-flip correlation function in Zubarev notation
\begin{eqnarray}
\label{eq:1dSpinFlipCorrFunc}
\langle \langle \delta B_{-p}; \delta B_p \rangle \rangle &=&
\frac{1}{N} \sum_{p} (1-2\chi)^2 \left \lbrace
\cos^2 \frac{p}{2} B^{(1,1)} (p,\omega) \right.
\nonumber  \\
&-& \cos \frac{p}{2} \sin \frac{p}{2}
\left (B^{(1,2)} (p,\omega) + B^{(2,1)} (p,\omega) \right )
\nonumber \\
 &+&  \left. \sin^2 \frac{p}{2}  B^{(2,2)} (p,\omega)
\right \rbrace
\end{eqnarray}
with particle-hole propagators
\begin{equation}
B^{(\mu,\nu)} (p,\omega) = \sum_{k,q} f_k^\mu f_q^\nu
\langle \langle c_k^\dagger c_{k+p} ;
c_{q+p}^\dagger c_q \rangle \rangle
\end{equation}
and the following form factors
\begin{equation}
f_k^0=1 \,\,\,\, , \,\,\,\, f_k^1=\cos k \,\,\,\, , \,\,\,\,
f_k^2=\sin k \,.
\end{equation}

\begin{figure}[t!]
\begin{center}
\epsfig{figure=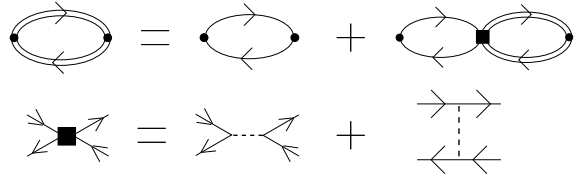,width=7.5cm,clip=}
\end{center}
\caption{Diagrammatic scheme for the extended RPA treatment
of Jordan-Wigner fermions in Eq.~(\ref{eq:RenPartHole1d}).}
\label{fig:RPA1d}
\end{figure}

Summing all particle-hole scattering processes, as illustrated in Fig.~\ref{fig:RPA1d} in
diagrammatic terms, 
a simple expression for the renormalized particle-hole 
propagator can be obtained
\begin{eqnarray}
{\rm B}^{(\mu,\nu)} (p,\omega) = {\rm b}^{(\mu,\nu)} (p,\omega)
&+& 2 J \cos p \, {\rm b}^{(\mu,0)} (p,\omega)  {\rm B}^{(0,\nu)} (p,\omega)
\nonumber \\
&-& 2 J \, {\rm b}^{(\mu,1)} (p,\omega) {\rm B}^{(1,\nu)} (p,\omega)
\nonumber \\
&-& 2 J \, {\rm b}^{(\mu,2)} (p,\omega) {\rm B}^{(2,\nu)} (p,\omega)
\, ,
\label{eq:RenPartHole1d}
\end{eqnarray}
where the noninteracting particle-hole propagators are given by
\begin{eqnarray}
{\rm b}^{(\mu,\nu)}(p,\omega) &=& \frac{1}{N} \sum_k f_k^\mu f_k^\nu \left \{
\frac{(1-\langle n_{k+p} \rangle ) \langle n_k \rangle}
{\omega + \epsilon_k - \epsilon_{p+k} + i 0^+}
\right. \\
&& \qquad\qquad\qquad - \left. \frac{(1-\langle n_k \rangle ) \langle n_{p+k} \rangle}
{\omega + \epsilon_k - \epsilon_{p+k} - i 0^+}
\right \} \,. \nonumber
\label{eq:NonInteractPartHole1d}
\end{eqnarray}

\begin{figure}[b!]
\begin{center}
\epsfig{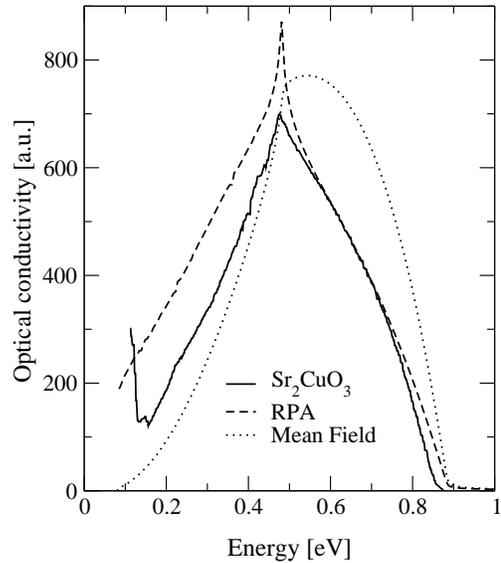}
\end{center}
\caption{$\sigma(\omega)$ obtained with
Jordan-Wigner fermions in comparison with the experimental 
optical conductivity of ${\rm Sr_2CuO_3}$ (solid
line), taken from Suzuura {\it et al.}~\cite{Suzuura}
Dotted line: mean-field
approximation, dashed line: RPA-approximation. Following
Lorenzana and Eder~\cite{LorenzanaEder} we 
have subtracted the same
linear background from the experimental data and we have used the 
same value for the exchange coupling
$J$=0.246eV and for the phonon-frequency $\omega_{ph}$=0.08eV.}
\label{fig:OptLeit1d}
\end{figure}

Evaluation of these equations 
determines $\sigma(\omega)$ which is shown in
Fig.~\ref{fig:OptLeit1d} in comparison with the experimental spectrum
of ${\rm Sr_2CuO_3}$ taken from Suzuura {\it et al.}~\cite{Suzuura}

A simple analysis of the experimental line shape of ${\rm Sr_2CuO_3}$
based on Jordan-Wigner fermions has already been discussed
by Suzuura {\it et al.}~\cite{Suzuura} in combination with the
experimental results. However, they restricted the evaluation to the
XY-model which corresponds to 
our mean field evaluation apart from a renormalization of the 
energy scale by a factor of $1+2/\pi$ in Eq.~(\ref{eq:1DMeanField}).
We find that it is
important to treat the two-particle correlation function $\sigma(\omega)$
at least within RPA. The resonance is shifted to lower energies
compared to the mean field approximation. In addition we observe a
cusp at $\omega=J(1+2/\pi)$ as a precursor of the logarithmic
singularity found by Lorenzana and Eder.~\cite{LorenzanaEder}
Despite the good agreement with respect
to the cusp and to the high energy side of the cusp, however, our RPA treatment
seems to overestimate the effect of the interaction because too
much spectral weight is shifted to energies below the cusp. This effect
could possibly be compensated by consideration of higher order 
correlations.

\section{Jordan-Wigner Transformation for the two-leg $S=\frac{1}{2}$-ladder}
\label{sec:JWladder}

Motivated by the results of our Jordan-Wigner fermion treatment for the
Heisenberg $S=\frac{1}{2}$--chain we ``slightly increase'' the dimensionality
and extend the approach to the nearest neighbor Heisenberg
two-leg $S=\frac{1}{2}$--ladder: 
\begin{equation}
\label{eq:HeisenbergLadder}
H = J_\perp \! \sum_i {\bf S}_{i,1} {\bf S}_{i,2}
  \!+ \! J \! \sum_i ({\bf S}_{i,1} {\bf S}_{i+1,1}
            \!+ {\bf S}_{i,2} {\bf S}_{i+1,2}) \, ,
\end{equation}
where $J_\perp$ is the exchange coupling along the rungs, $J$ the
coupling along the legs, $i$ refers to the site index along the legs
and the subscripts $1,2$ label the two different legs.

\begin{figure}[t!]
\begin{center}
\epsfig{figure=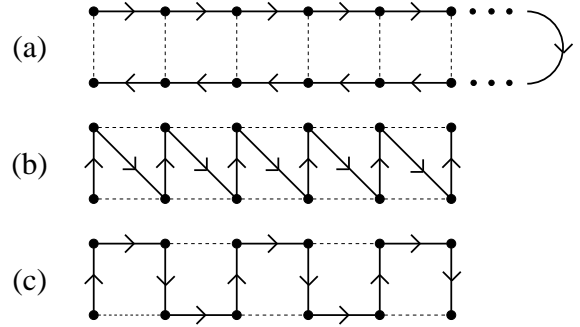,width=7.5cm,clip=}
\end{center}
\caption{Possible path configurations for a two-leg ladder.}
\label{fig:JordanWignerPfade}
\end{figure}

Generalizations of the Jordan-Wigner transformation 
to higher dimensions have been suggested~\cite{JW2d,JW3d}
and may be adopted for spin ladders.
For spin ladders the one-dimensional Jordan-Wigner transformation can
be applied directly, when all spins are arranged in a one-dimensional sequence. 
With this scheme we can control the range of the
interaction terms through a convenient choice of a path covering all
sites, whereas the application of a two-dimensional representation
to the spin ladders would generate
long-range interaction terms in the Hamiltonian.
Possible path configurations
through a two-leg ladder are shown in Fig.~\ref{fig:JordanWignerPfade}.

The path displayed in Fig.~\ref{fig:JordanWignerPfade}(a) is obviously
very close to the one-dimensional situation. As a consequence
the rung interaction is difficult to treat in this
representation because every product of neighboring rung-spins
contains an infinite number of phase-factors. 
The rung-coupling, however, is a relevant perturbation
since the excitation spectrum of a two-leg ladder remains gapped for
all coupling ratios $J_\perp/J$. 
Therefore a path which passes through all the rungs should be more suitable.
Possible realizations are a zigzag path\cite{Azzouz} and a
meander path\cite{DaiSu}, displayed in
Fig.~\ref{fig:JordanWignerPfade}(b) and (c), respectively. Although 
the zigzag path  
appears simpler and more symmetric at first sight,
we show in Appendix~\ref{app:DimLimit} that 
within a mean field treatment (analogous to the previous section)
only the meander path can reproduce the correct strong coupling limit
of the one-triplet dispersion $\epsilon_k=J_\perp +J \cos k$
for $J_\perp/J \gg 1$. Therefore we  chose the meander path
even though the nearest neighbor
Heisenberg-Hamiltonian is slightly more complicated in this representation.

\begin{figure}[t!]
\begin{center}
\epsfig{figure=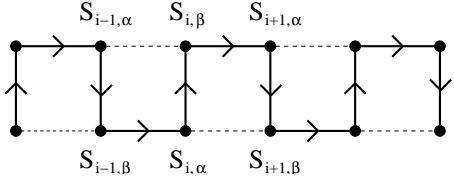,width=6cm,clip=}
\end{center}
\caption{Sublattice structure for the meander path.}
\label{fig:Sublattice}
\end{figure}

Following Dai and Su~\cite{DaiSu} we divide the ladder into two
sublattices as indicated in Fig.~\ref{fig:Sublattice}.
Introducing two species of spinless fermions $\alpha_i$ and $\beta_i$,
the spin operators on the two sublattices transform as:
\begin{eqnarray}
\label{eq:JWtrafo}
S_{i,\alpha}^+ &=& \alpha_i^\dag e^{i\pi\sum_{j<i}
         (\alpha_j^\dag \alpha_j + \beta_j^\dag \beta_j)}
\nonumber \\
S_{i,\beta}^+ &=& \beta_i^\dag e^{i\pi\sum_{j<i}
      (\alpha_j^\dag \alpha_j + \beta_j^\dag \beta_j)}
                       e^{i\pi \alpha_i^\dag \alpha_i} \, ,
\end{eqnarray}
where the summation in the phase factor is along the meander path.
For products of spin operators, which are not successive along the
meander path,
like e.g.\ $S_{i,\alpha}^+S_{i+1,\beta}^-+S_{i,\alpha}^-S_{i+1,\beta}^+$,
the phases corresponding to intermediate sites along the meander
path do not cancel. This is different from the one-dimensional situation
where all nearest-neighbor spin operators are also successive along
the path. Using transformation~(\ref{eq:JWtrafo}) the Heisenberg Hamiltonian of
Eq.~(\ref{eq:HeisenbergLadder}) becomes: 
\begin{eqnarray}
\label{eq:HeisenbergLadderJW}
H = &J_\perp& \sum_i \left \{
\frac{1}{2} (\alpha_i^\dagger \beta_i +\beta_i^\dagger \alpha_i)
+(\alpha_i^\dagger\alpha_i-\frac{1}{2})(\beta_i^\dagger\beta_i-\frac{1}{2}) 
 \right \}  \nonumber \\
+ &J& \sum_i \left \lbrace \,
 \frac{1}{2} \left [ \beta_i^\dagger \alpha_{i+1} 
+\alpha_i^\dagger\beta_{i+1} e^{i\pi(n_{\beta_i}+n_{\alpha_{i+1}})} 
+ {\rm H.c.} \right ]  \right. \nonumber \\
& &  \qquad\quad + \,(\alpha_i^\dagger\alpha_i-\frac{1}{2})
             (\beta_{i+1}^\dagger\beta_{i+1}-\frac{1}{2})
\nonumber \\
& & \left. \qquad\quad + \,(\alpha_{i+1}^\dagger\alpha_{i+1}-\frac{1}{2})
       (\beta_i^\dagger\beta_i-\frac{1}{2})
    \right \} \, . 
\end{eqnarray}
Unfortunately the phase factor
$e^{i\pi(n_{\beta_i}+n_{\alpha_{i+1}})}$ 
from spin products of non-successive sites cannot be treated exactly.
Dai and Su have 
replaced this phase factor by its average value. 
This treatment, however, can be improved in a systematic way by
expanding the phase factor:
\begin{equation}
e^{i\pi(n_{\beta_i}+n_{\alpha_{i+1}})}=(1-2\beta_i^\dagger\beta_i)
(1-2\alpha_{i+1}^\dagger \alpha_{i+1}) 
\end{equation}
and reinserting this expansion into Hamiltonian
(\ref{eq:HeisenbergLadderJW}). In this way we obtain additional
interaction terms containing 4-- and 6--fermion operators
which we now treat on the same footing as the
Ising-interaction terms.

\subsection{Mean field treatment -- Ladder}
\label{sec:MeanFieldLadder}

Inspired by the results of the mean-field treatment 
for the spin chain, we adopt a similar approach here and
consider all possible nearest-neighbor bond amplitudes:
\begin{equation}
\label{eq:bondamp}
\chi_0=\langle \beta_i^\dagger \alpha_i \rangle \,\,\, , \,\,\,
\chi_1=\langle \beta_i^\dagger \alpha_{i+1} \rangle \,\,\, , \,\,\,
\chi_2=\langle \alpha_i^\dagger \beta_{i+1} \rangle \,.
\end{equation}
Taking into account all possible contractions of the 4- and
6-fermion operator terms we arrive at the following mean-field Hamiltonian:
\begin{equation}
H_{MF}=\sum_k (\gamma_k \alpha_k^\dagger \beta_k + H.c.)
\label{eq:HMF}
\end{equation}
with
\begin{eqnarray}
\label{eq:gamma}
\gamma_k &=& J_\perp (\frac{1}{2}-\chi_0)
     + 4 J \chi_0\chi_1 \\   
     & & + \,\, J \cos k (\frac{1}{2}+2\chi_0^2-4\chi_1\chi_2-\chi_1-\chi_2-2\chi_1^2)   
	 \nonumber \\
     & & + \,\, i J \sin k
         (\frac{1}{2}+2\chi_0^2-4\chi_1\chi_2-\chi_1+\chi_2+2\chi_1^2)
     \,.
         \nonumber
\end{eqnarray}
This expression has already been simplified using that
$\chi_0$,  $\chi_1$ and  $\chi_2$  turn out to be real.
The above Hamiltonian can easily be diagonalized
\begin{equation}
\label{eq:Hdiag}
H_{MF}=\sum_k \epsilon_k (\tilde \alpha_k^\dagger \tilde \alpha_k
                         -\tilde \beta_k^\dagger \tilde \beta_k)
\quad \textrm{with} \quad \epsilon_k=\vert \gamma_k \vert 
\end{equation}
using
\begin{eqnarray}
\label{eq:diagtrafo}
\alpha_k = \frac{1}{\sqrt{2}} u_k (\tilde \alpha_k + \tilde \beta_k)
\quad &,& \quad
\beta_k = \frac{1}{\sqrt{2}} v_k (\tilde \alpha_k - \tilde \beta_k)
\nonumber \\
\qquad u_k=v_k^*=e^{i\phi_k/2} \quad &,& \quad
\gamma_k=\vert \gamma_k \vert e^{i\phi_k} \, .
\end{eqnarray}

\begin{figure}[t!]
\begin{center}
\epsfig{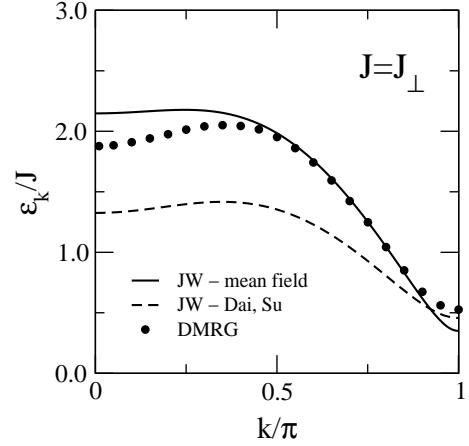}
\end{center}
\caption{Dispersion for the isotropic ladder $J=J_\perp$.
Solid line: mean field dispersion for Jordan-Wigner fermions; 
dashed line: dispersion obtained by averaging the
phase factor analogous to the treatment by Dai and Su;\cite{DaiSu}
circles: one-triplet dispersion 
obtained with DMRG for a $N$=80--site ladder.\cite{condmat}}
\label{fig:DispersionMFJ10}  
\end{figure}

\begin{figure}[b!]
\begin{center}
\epsfig{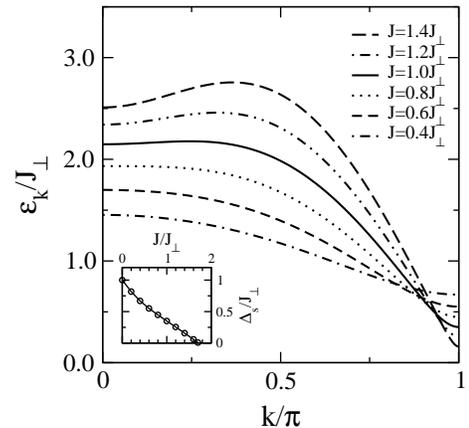}
\end{center}
\caption{Mean-field dispersion for Jordan-Wigner fermions 
for different coupling ratios $J/J_\perp$. Inset: spin gap
$\Delta_s$ as a function of $J/J_\perp$.}
\label{fig:DispersionGapMF}  
\end{figure}

For the isotropic ladder, i.e.\ $J=J_\perp$, we obtain $\chi_0=-0.3617$,
$\chi_1=-0.2679$ and $\chi_2=0.1777$. 
As the product of the bond amplitudes around a plaquette is negative,
our mean field treatment corresponds to a $\pi$-flux
state of the spinless fermions. 
Note, that this $\pi$-flux phase is different from that of
Azzouz {\it et al.},~\cite{Azzouz} who have by construction replaced
the complete phase factor by a $\pi$-flux phase. Both approaches are
compared in Appendix~\ref{app:PhaseFactor}.

The resulting dispersion is
displayed in Fig.~\ref{fig:DispersionMFJ10}, in comparison with the
dispersion for a $N$=80-site ladder obtained by DMRG.\cite{condmat}
For momenta between $k \approx 0.5\pi$--$0.9\pi$   we find
nearly perfect agreement with the DMRG results. The spin gap
at $k=\pi$, however, is too small and for momenta 
$k < \pi/2$ the energy is overestimated. Nevertheless, our mean-field 
treatment is able to reproduce a dip for small momenta,
which is a precursor of the symmetric (with respect to $k=\pi/2$)
spinon dispersion of the spin chain. The dip becomes more pronounced 
when the leg coupling $J/J_\perp$ is increased as can be seen in
Fig.~\ref{fig:DispersionGapMF}. Note that within a mean-field
treatment of the bosonic bond-operator representation of  
elementary rung-triplets~\cite{GopalanRice,SushkovKotov} it has not been
possible to achieve a dispersion-dip for the isotropic ladder.

To demonstrate the improvement of our mean-field evaluation with
respect to the mean-field treatment by Dai and Su,~\cite{DaiSu}
who have 
replaced the phase factor by its expectation value,
we have added their dispersion in Fig.~\ref{fig:DispersionMFJ10}.
Qualitatively it is very similar to our mean-field dispersion.
Its magnitude, however, is by a factor of about 1.5 too small over
a large part of the Brillouin zone. Therefore we conclude that
an adequate treatment of the phase factor is very important and it is 
necessary to go beyond  
averaging of the phase factor.

The spin gap $\Delta_s$, which is shown in the inset of
Fig.~\ref{fig:DispersionGapMF} as a function of $J/J_\perp$, is
underestimated by our approach. It even vanishes for $J \approx 1.7
J_\perp$. As the spin gap is known to be finite for all finite
$J/J_\perp$, this coupling ratio marks the breakdown of our mean field
treatment.

\subsection{Extended RPA-treatment -- Ladder}
\label{sec:RPALadder}

The optical conductivity $\sigma(\omega)$ constitutes a powerful probe of the
spin excitations of a spin ladder. 
It has been possible to verify experimentally the existence of a 
$S$=0--two triplet bound state  in the optical conductivity of the
spin ladder compound (La,Ca)$_{14}$Cu$_{24}$O$_{41}$\cite{PRL2001}
and a more refined analysis has shown
that it is necessary to include a 4-spin cyclic exchange interaction~\cite{condmat} 
of about $J_{cyc} \approx 0.20-0.27 J_\perp$.
Bound states in the singlet and triplet excitation channel were
predicted by Sushkov and Kotov~\cite{SushkovKotov} and by Damle and 
Sachdev,\cite{damle} similar to the states analyzed by Uhrig and 
Schulz~\cite{uhrigschulz} in dimerized spin chains.
More extensive perturbative investigations, within a linked cluster expansion, were
performed by Trebst {\it et al.}~\cite{monienPRL} and 
the observability of the singlet bound state was suggested
by Jurecka and Brenig\cite{jurecka}.

Here, we will demonstrate, how the spin-flip correlation
functions, which contribute to $\sigma(\omega)$, can be obtained within
the Jordan-Wigner approach. We discuss the resultant correlation
functions for an isotropic ladder $J=J_\perp$,
without cyclic exchange ($J_{cyc}=0$), 
and focus on the $S$=0--bound state and the continuum excitations.
For the calculation of the optical conductivity we will concentrate on an
isolated ${\rm Cu_2O_3}$-ladder. The phonon-assisted magnetic
contribution to $\sigma(\omega)$ results largely from the
simultaneous excitation of two neighboring spin-flips plus
a Cu-O bond-stretching phonon:
\begin{equation}
\sigma( \omega) \sim  - \omega \sum_{p} \sum_{p_y=0,\pi}
          f_p \, {\rm Im} \langle \langle \delta B_{-{\bf p}} ; \delta B_{\bf p}
                  \rangle \rangle_{( \omega- \omega_{ph})}
\label{eq:sigma}
\end{equation}
where ${\bf p}=(p,p_y)$ and the operators
\begin{eqnarray}
\label{eq:SpinflipOp}
\delta B_{\bf p}^{\rm leg}  &=& \frac{1}{N} \! \sum_i \!
    \sum_{l=1\!,2} e^{i{\bf p}{\bf r}_{i\!,l}} \! \left (
      {\bf S}_{i,l} {\bf S}_{i+1\!,l}
        \! - \! \langle {\bf S}_{i,l} {\bf S}_{i+1\!,l} \rangle
      \right )
\nonumber \\
\delta B_{\bf p}^{\rm rung} &=& \frac{1}{N} \sum_i e^{i{\bf p}{\bf r}_i} \left (
      {\bf S}_{i,1} {\bf S}_{i,2}
        - \langle {\bf S}_{i,1} {\bf S}_{i,2} \rangle \right )
\end{eqnarray}
are the spin-flip operators for polarization of the electrical field along
the legs and the rungs, respectively.

Following our previous treatment in Ref.~\onlinecite{condmat}, we consider
phonon form factors given by:
\begin{equation}
\label{eq:formfactor}
f_p^{\rm leg} = 8 \sin^4 (\frac{p}{2}) \,\,\,\, ,\,\,\,\,
f_p^{\rm rung} = 8 \sin^2 (\frac{p}{2}) + 4 \,.
\end{equation}
Here, $f_p^{\rm leg}$ originates from the coupling to in-phase and
out-of-phase stretching modes of   
O-ions on the legs and is the same form factor as for an isolated spin chain.
For $f_p^{\rm rung}$  we take in addition to the 
out-of-phase stretching mode also the vibration of the
O-ion on the rung into account,
which is responsible for the constant contribution in Eq.~(\ref{eq:formfactor}).

For the calculation of the spin-flip correlation function we apply the 
Jordan-Wigner transformation~(\ref{eq:JWtrafo}) to the spin-flip
operators ${\bf S}_{i,1} {\bf S}_{i+1,1} \pm {\bf S}_{i,2} {\bf
S}_{i+1,2}$ and ${\bf S}_{i,1} {\bf S}_{i,2}$.
All terms with 4 and 6 fermion operators are reduced to
two-operator terms by replacing all surplus operators with their
contractions~(\ref{eq:bondamp}). 
With this procedure the Fourier transform of the
spin-flip operators becomes:   
\begin{eqnarray}
\label{eq:SpinflipOpMF}
\delta B_p^{\rm rung} &=& \frac{1}{\sqrt{N}} \sum_k (\frac{1}{2}-\chi_0)
\left ( \alpha_k^\dag \beta_{p+k} + \beta_k^\dag \alpha_{p+k} \right )
\\
\delta B_{p,p_y=0}^{\rm leg} &=& \frac{1}{\sqrt{N}}  \sum_k
\left \{ \alpha_k^\dag \beta_{p+k} (a + be^{ik} - ce^{-i(p+k)})
\right. \nonumber \\
    & & \qquad\quad\,\,  + \left. \beta_k^\dag \alpha_{p+k} (a + be^{-i(p+k)} - ce^{ik})
\right \}
\nonumber \\
\delta B_{p,p_y=\pi}^{\rm leg} &=& \frac{1}{\sqrt{N}}  \sum_k
\left \{ \alpha_k^\dag \beta_{p+k+\pi} (-a + \tilde b e^{ik} - c e^{-i(p+k)})
\right. \nonumber \\
    & & \qquad\quad\,\,  + \left. \beta_k^\dag \alpha_{p+k+\pi}
         (-a - \tilde be^{-i(p+k)} + ce^{ik}) \right \}
\nonumber 
\end{eqnarray}
with
\begin{eqnarray}
a=4\chi_0\chi_1 \,\, , \,\,
b=\frac{1}{2}+2\chi_0^2-4\chi_1\chi_2-\chi_1
\nonumber \\
c=2\chi_1^2+\chi_2 \,\, , \,\,
\tilde b =\frac{1}{2}-2\chi_0^2+4\chi_1\chi_2-\chi_1 \,.
\end{eqnarray}

Inserting the spin-flip operators~(\ref{eq:SpinflipOpMF}) into the
optical conductivity~(\ref{eq:sigma}) produces a sum of particle-hole
propagators with different form factors. To evaluate these particle-hole
propagators in RPA we prefer to use the original fermionic operators
$\alpha, \beta$ because transformation to the operators $\tilde \alpha,
\tilde \beta$~(\ref{eq:diagtrafo}) would increase
the number of interaction terms considerably.

\begin{figure}[t!]
\begin{center}
\epsfig{figure=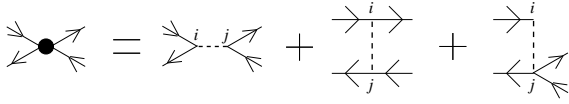,width=7.5cm,clip=}
\end{center}
\caption{Diagrammatic representation of the processes which contribute to
the extended RPA-treatment, in real space.}  
\label{fig:RPAvertex}  
\end{figure}

Prior to the derivation of the RPA-equations, the interaction terms
in the Hamiltonian have to be reduced to two-particle interactions
in order to deal only with 4--particle vertices.
Accordingly,
all 6--operator terms, which appear in Eq.~(\ref{eq:HeisenbergLadderJW}), are
reduced to 4--operator terms by replacing all possible contractions
with the corresponding bond amplitudes~(\ref{eq:bondamp}). 
In this way, we obtain the following reduced interaction term from
Hamiltonian~(\ref{eq:HeisenbergLadderJW}):
\begin{eqnarray}
\label{eq:HRPA}
&H_{\rm red}& = J \frac{1}{N} \sum_{k_1...k_4} \delta(k_1+k_2-k_3-k_4)
\nonumber \\
& & \!\!\!\!\!\!\!  \left \{
  \alpha_{k_1}^\dag \beta_{k_2}^\dag \beta_{k_3} \alpha_{k_4}
    \left [ \frac{J_\perp}{J} + e^{i(k_2-k_3)}
            + (1+4\chi_2) e^{-i(k_2-k_3)} \right. \right.
\nonumber \\
& & \qquad\qquad\qquad\qquad\!  +\, 2 \chi_1 (e^{i(k_1+k_2)} + e^{-i(k_1+k_2)}) 
\nonumber \\
 & &  \qquad \qquad \qquad\qquad\! \left.
    -\, 2\chi_0 (e^{ik_1} + e^{-ik_2} + e^{ik_3} + e^{-ik_4})
    \right ]
\nonumber \\
 & & \!\!\!\!\!\!\!  + \, \alpha_{k_1}^\dag \alpha_{k_2}^\dag \beta_{k_3} \beta_{k_4}
      \,\, 2 \chi_1 e^{i(k_2-k_3)}
\nonumber \\
 & & \!\!\!\!\!\!\! \left. + \, \beta_{k_1}^\dag \beta_{k_2}^\dag \alpha_{k_3} \alpha_{k_4}
      \,\, 2 \chi_1 e^{i(k_1-k_4)} \right \}
\end{eqnarray}
A set of RPA-equations for the particle-hole propagators,
which are listed in Appendix~\ref{app:RPAequations}, can be obtained
by consideration of all possible vertex-configurations of the
interaction Hamiltonian~(\ref{eq:HRPA}). 
In real space these vertices correspond not only to a summation of
bubble-diagrams, but also include ladder-diagrams and
other non-local terms as indicated in
Fig.~\ref{fig:RPAvertex}.
For this reason we use the term {\it extended} RPA-treatment.

\subsection{Spin-flip correlation functions}
\label{sec:SpinFlipCorrFuncLadder}

\begin{figure}[b!]
\begin{center}
\epsfig{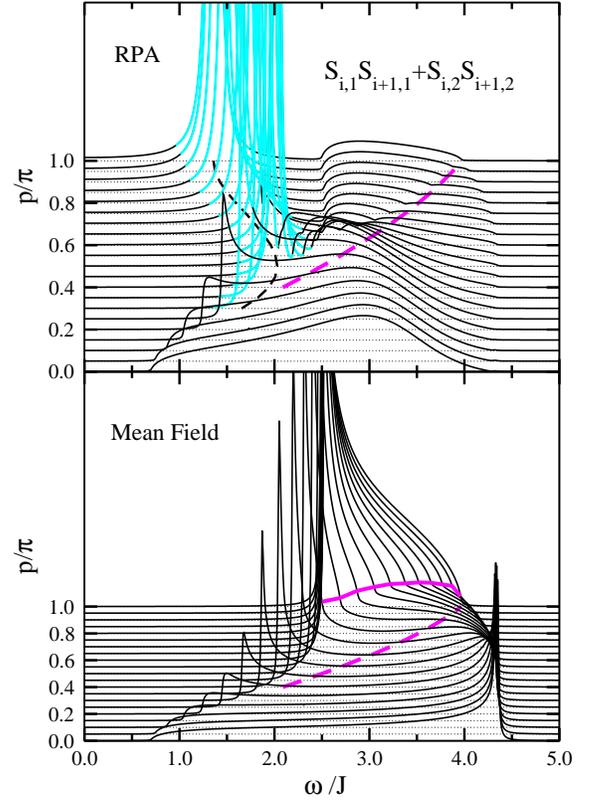}
\end{center}
\caption{RPA and mean-field evaluation of the momentum resolved
correlation function $\langle \langle \delta B_{-p,p_y=0}^{\rm leg}; \delta
B_{p,p_y=0}^{\rm leg} \rangle \rangle$, where a broadening of $\delta=0.01
J$ has been used. The grey lines 
(upper panel) designate the $S$=0-bound state.
The dashed dark grey line is a projection of the dark grey line in the
lower panel, which links the points of sharp increase in the middle of
the continuum. This is a precursor of the upper edge of the 2-spinon
continuum in single chains.~\cite{PRLKoeln,MarkusExpTheo}
In RPA a dip structure remains at the same position
in the continuum.}
\label{fig:JW_p0}  
\end{figure}

\begin{figure}[b!]
\begin{center}
\epsfig{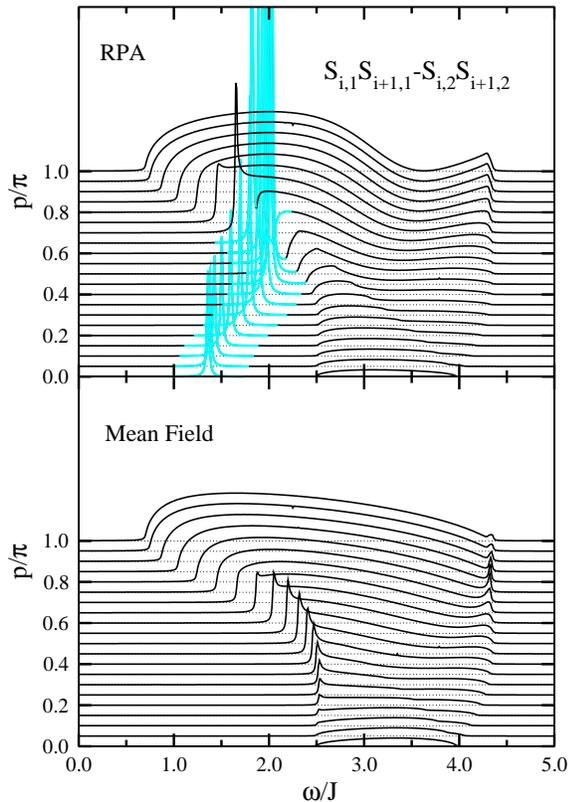}
\end{center}
\caption{RPA and Mean-field evaluation of the correlation function
$\langle \langle \delta B_{-p,p_y=\pi}^{\rm leg}; \delta
B_{p,p_y=\pi}^{\rm leg} \rangle \rangle$, where a broadening of $\delta=0.01
J$ has been used. The grey lines indicate the $S$=0-bound state.}
\label{fig:JW_ppi}  
\end{figure}

\begin{figure}[t!]
\begin{center}
\epsfig{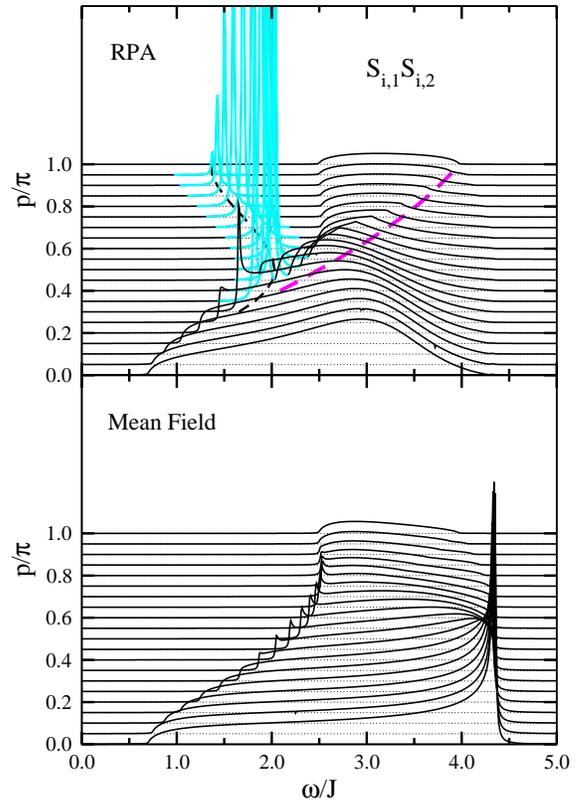}
\end{center}
\caption{RPA and Mean-field evaluation of the correlation function
$\langle \langle \delta B_{-p}^{\rm rung}; \delta
B_{p}^{\rm rung} \rangle \rangle$, where a broadening of $\delta=0.01
J$ has been used. The grey lines indicate the $S$=0-bound state.
The dashed dark grey line in the upper panel is the same as in
Fig.~\ref{fig:JW_p0}. 
}
\label{fig:JW_rung}  
\end{figure}

We have calculated the correlation functions for spin-flips along the legs
$\delta B_{p,p_y=0}^{\rm leg}$, $\delta B_{p,p_y=\pi}^{\rm leg}$ and
for spin-flips along the rungs $\delta B_p^{\rm rung}$
for an isotropic ladder $J=J_\perp$ using the
RPA-treatment of the previous section. The results are 
presented in Figs.~\ref{fig:JW_p0}, \ref{fig:JW_ppi} and \ref{fig:JW_rung},
respectively. For comparison the mean-field evaluation of each of the
correlation functions is displayed in the lower panels.

\begin{figure}[t!]
\begin{center}\epsfig{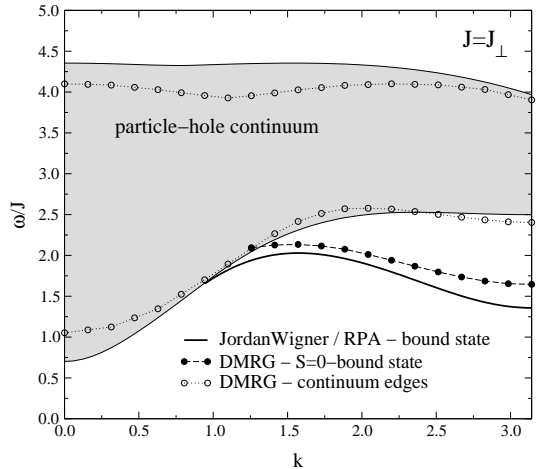}
\end{center}
\caption{Particle--hole continuum of the Jordan--Wigner fermions and
$S$=0--bound state (thick solid line) in comparison with 
the two-triplet continuum (open symbols) and $S$=0--bound state (filled symbols)
obtained by DMRG for a ladder with $N$=80--sites.\cite{condmat}}
\label{fig:BoundState}  
\end{figure}

In the mean-field evaluation of
$\langle \langle \delta B_{-p,p_y=0}^{\rm leg};
\delta B_{p,p_y=0}^{\rm leg} \rangle \rangle$
(lower panel of Fig.~\ref{fig:JW_p0}) one observes van-Hove
singularities at the upper edge of the continuum for small momenta and
at the lower edge of the continuum for large momenta. 
With the RPA-treatment the
van-Hove singularities at the continuum edges disappear. For small
momenta the maximum of the continuum is shifted 
from the upper edge downwards to about $\omega \approx 3 J$.
At large momenta we see the formation of the $S$=0-bound
state. The bound state emerges from the continuum 
at approximately $k \approx 0.3 \pi$, it passes through a maximum at 
$k \approx \pi/2$ and a minimum at $k = \pi$. In
Fig.~\ref{fig:BoundState} the dispersion of the bound state is
compared with the DMRG calculation for a $N$=80--site ladder.~\cite{condmat} We
find good agreement between both methods, only the energy of the RPA-dispersion
is slightly too low. This indicates that the interaction strength is 
somewhat overestimated in RPA.

In the mean field evaluation of the out-of-phase component of the
correlation function for spin-flips along the legs $\langle \langle
\delta B_{-p,p_y=\pi}^{\rm leg}; \delta B_{p,p_y=\pi}^{\rm leg}
\rangle \rangle$ (lower panel in Fig.~\ref{fig:JW_ppi}) the van--Hove
singularities at the continuum edges are suppressed. 
The overall momentum dependence on $p$ appears to be reversed when
the out-of-phase component in Fig.~\ref{fig:JW_ppi} (upper panel) is compared to the
in-phase component in Fig.~\ref{fig:JW_p0} (upper panel).
This is caused by the checker-board sublattice structure of the meander
path, which shifts the momentum of the particle-hole propagator 
$\delta B_{p,p_y=\pi}^{\rm leg}$ by $\pi$ in relation~(\ref{eq:SpinflipOpMF}).
The inversion of the momentum dependence is especially noticeable for the bound 
state. However, the out-of-phase component should not contain 
the bound state at all but only contribute to the continuum excitations,
an issue that we have addressed previously in Ref.~\onlinecite{condmat}.
The argument is based on the observation that the out-of-phase component originates
from the excitation of 3 different rung-triplets,~\cite{condmat} 
when it is expressed in terms of rung-triplet operators.~\cite{SachdevBhatt,GopalanRice}  
The $S$=0--bound state, on the other hand, arises from
scattering processes of two equal triplets and therefore cannot be
present in the out--of--phase component.  Although the spectral weight of
the bound state in the $p_y=\pi$ component is only small, this
indicates a failure of our approach. This failure is probably related
to the fact that the meander path breaks certain symmetries of the
original ladder model.  In our mean-field evaluation we find,
e.g., $\langle {\bf S}_{i,1} {\bf S}_{i+1,1} \rangle \ne \langle {\bf
S}_{i,2} {\bf S}_{i+1,2 \rangle}$.  This artificial symmetry breaking
will be discussed in more detail in Appendix~\ref{app:SymmetryBreaking}.
However, the resultant effects may average out to some extent in the sum of the
spin-flip operators on both legs ${\bf S}_{i,1} {\bf S}_{i+1,1}+{\bf
S}_{i,2} {\bf S}_{i+1,2}$, whereas they will probably even be
enhanced in the difference ${\bf S}_{i,1} {\bf S}_{i+1,1}-{\bf
S}_{i,2} {\bf S}_{i+1,2}$. This may explain the 
shortcomings of our approach with respect to the out-of-phase component, 
while we find reasonable results for the in-phase-component.

In Fig.~\ref{fig:JW_rung} the RPA- and mean-field evaluation for the
correlation function of spin-flips along the rungs $\langle \langle
\delta B_{-p}^{\rm rung}; \delta B_{p}^{\rm rung} \rangle \rangle$ are
shown in the upper and lower panel, respectively. For $p=0$ it is
identical to the $p=0$--component of the in-phase
correlation function for spin-flips along the legs $\langle \langle
\delta B_{-p,p_y=0}^{\rm leg}; \delta B_{p,p_y=0}^{\rm leg} \rangle
\rangle$ and both correspond to the correlation function for the Raman
response.~\cite{FreitasSingh,RamanKoeln}  For larger momenta, the
spectral weight of the rung-correlation is much smaller than the
in-phase component of the leg-correlations and at $p=\pi$ 
the weight of the $S$=0 bound state vanishes according to a 
selection rule.~\cite{PRL2001}

\subsection{Optical conductivity}
\label{sec:SigmaLadder}

\begin{figure}[b!]
\begin{center}
\epsfig{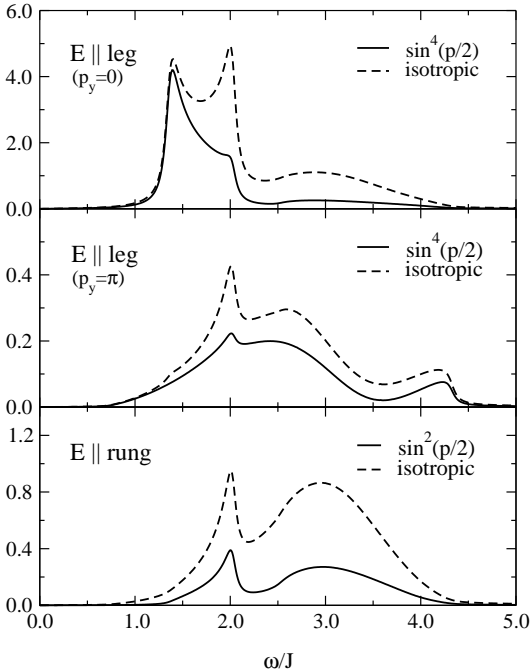}
\end{center}
\caption{Momentum integrated spin--flip correlation functions, which
contribute to $\sigma(\omega)$ for polarization along the legs (top
and middle panel) and for polarization along the rungs (bottom
panel). In order to visualize the resulting contribution to
$\sigma(\omega)$ in Eq.~(\protect{\ref{eq:sigma}}) each correlation
function has been multiplied with 
the frequency $\omega$ and for each of them the momentum
integration without prefactor (dashed lines) and with the corresponding
prefactors of Eq.~(\protect{\ref{eq:formfactor}}) (solid lines) are shown.}
\label{fig:sigmaJW}  
\end{figure}

Once the momentum dependent spin-flip correlation functions,
shown in Figs.~\ref{fig:JW_p0}, \ref{fig:JW_ppi} and
\ref{fig:JW_rung}, are known the optical conductivity $\sigma(\omega)$
in Eq.~(\ref{eq:sigma}) can easily be obtained by integration.
In  Fig.~\ref{fig:sigmaJW} the momentum integrated spin-flip
correlation functions are
displayed whereby we present them with form factors equal to unity
(dashed lines) and with the form factors of Eq.~(\ref{eq:formfactor}) 
(solid lines).
To demonstrate their contribution to $\sigma(\omega)$, all spectra 
have already been multiplied with frequency $\omega$.

For polarization parallel to the legs and $p_y=0$ (top panel of
Fig.~\ref{fig:sigmaJW}) two dominant
peaks appear at $\omega_1 \approx 1.4 J$ and $\omega_2 \approx 2.0 J$.
They are caused by van Hove singularities arising from the dispersion of
$S$=0--bound state~\cite{PRL2001} at $p=\pi$ and $p \approx \pi/2$. 
The upper peak at $\omega_2$ is
suppressed by consideration of the relevant form factor
$\sin^4(p/2)$. Also the continuum excitations are reduced 
considerably because they originate largely from small
momenta, where no bound state is present, see Fig.~\ref{fig:JW_p0}. The
out-of-phase component (middle panel of Fig.~\ref{fig:sigmaJW})  is an order
of magnitude smaller than the in-phase component. We consider
the results for the out-of-phase component, however, less reliable due to
symmetry breaking by the meander path,
as mentioned in the previous section.

For polarization parallel to the rungs (bottom panel of
Fig.~\ref{fig:sigmaJW}) two form factors contribute to
$\sigma(\omega)$, a constant and a $\sin^2(p/2)$--term. Only the
upper bound state at $\omega_2$ is present in the momentum integrated
spectrum because the $S$=0--bound state is suppressed at $p=\pi$ 
in accord with a selection rule,~\cite{PRL2001} see also Fig.~\ref{fig:JW_rung}.

\begin{figure}[t!]
\begin{center}\epsfig{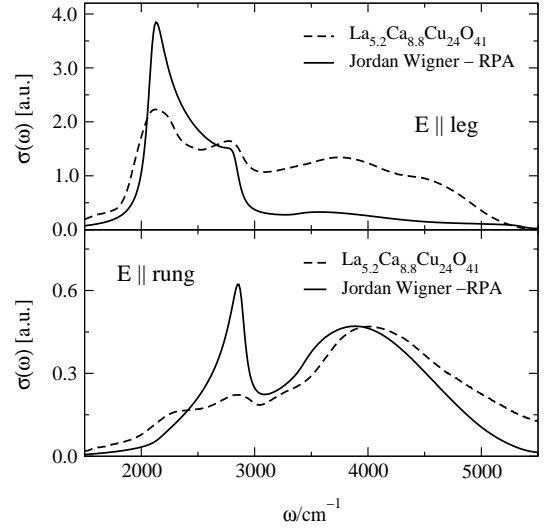}
\end{center}
\caption{Comparison of the Jordan--Wigner spectra for the isotropic
ladder using an exchange coupling of $J$=1100cm$^-1$
with the experimental $\sigma(\omega)$ of
${\rm La_{5.2}Ca_{8.8}Cu_{24}O_{41}}$. The experimental spectra
have been measured by Windt and Gr\"uninger.~\cite{condmat}
For the phonon a frequency of
$\omega_{ph}$=600cm$^{-1}$ has been assumed for polarization along
the legs and $\omega_{ph}$=650$^{-1}$ for polarization along the rungs.}
\label{fig:sigmaJWexp}  
\end{figure}

In Fig.~\ref{fig:sigmaJWexp} our spectra are compared with the
experimental optical conductivity of ${\rm La_{5.2}Ca_{8.8}Cu_{24}O_{41}}$
for polarization along the legs (top panel) and for polarization along
the rungs (bottom panel). Following our previous treatment in
Ref.~\onlinecite{PRL2001}, an exchange coupling of $J$=1100cm$^{-1}$,
a phonon frequency of $\omega_{ph}$=600cm$^{-1}$ for polarization along
the legs and $\omega_{ph}$=650cm$^{-1}$ for polarization along the rungs
have been used. The difference of 50cm$^{-1}$ in the phonon frequencies for
polarization along the legs and the rungs accounts for the
shift observed experimentally with respect to the upper bound state in both
polarizations.~\cite{PRL2001,condmat}

For polarization along the legs we find
good agreement with respect to the bound states. The continuum
contribution, however, is strongly underestimated. 
To some extent this is due to the fact that, 
so far, we considered only the creation of two fermions at the
external current vertex: we  approximated the spin-flip operators
in Eq.~(\ref{eq:SpinflipOpMF}) by a particle-hole creation operator.
Without this approximation the correlation functions $\langle \langle
\delta B_{-p,p_y=0,\pi}^{\rm leg}; \delta B_{p,p_y=0,\pi}^{\rm leg}
\rangle \rangle$ would contain also the
excitation of 4-- and 6--fermions, which would increase the
amount of high-energy excitations. The weight of these higher
order excitations will be estimated in the next section.

For polarization parallel to the rungs we find very good agreement
even with respect to the weight and line shape of the continuum. 
Rung correlations are obviously treated quite accurately in our meander-path
representation. There are no phase factors in the product of two spin
operators on the same rung because they are neighboring along the
meander path. Therefore the only 4--fermion process results from 
the Ising-like term in the rung operator~(\ref{eq:SpinflipOp}).

\subsection{Higher order contributions}
\label{sec:HigherOrderContributions}

The contribution of 4-- and 6--fermions to the spin-flip correlation
functions in leg polarization can be obtained analogous to
the 2--fermion contribution. In the spin-flip operator
$\delta B_p^{\rm leg}$, Eq.~(\ref{eq:SpinflipOp}), the spin operators
are replaced by fermionic operators. Terms with 6 operators are designated as the 
6--fermion part. Terms with 4 and 6 operators contribute to
the 4--fermion part, when all surplus operators are replaced by their
contractions, analogous to Eq.~(\ref{eq:SpinflipOpMF}).  
With this scheme one obtains for the 4-- and 6--fermion part of the
spin-flip operator
\begin{eqnarray}
& & \delta B_{p,p_y=0}^{\rm 4,leg} = \frac{1}{N^{3/2}}
\sum_{k_1 \ldots k_4}
2 \, \delta(p \!+\! k_1 \!+\! k_2 \!-\! k_3 \!-\! k_4) \times
\\
& & \quad \left \{ \alpha_{k_1}^\dag \beta_{k_2}^\dag \beta_{k_3} \alpha_{k_4}
\left [ \chi_1 (e^{i(k_1+k_2)}+e^{-i(k_3+k_4)}) \right.\right.
\nonumber \\
& & \qquad\qquad\qquad\quad -\chi_0
(e^{ik_1}+e^{-ik_4}+e^{-i(p+k_2)}+e^{-i(p-k_3)})
\nonumber \\
& & \qquad\qquad\qquad\quad  +\frac{1}{2} \left. (e^{i(k_2-k_3)}+e^{i(k_1-k_4)})
+2 \chi_2 e^{i(k_1-k_4)} \right ]
\nonumber \\
& & \quad \left. + \alpha_{k_1}^\dag \alpha_{k_2}^\dag \beta_{k_3} \beta_{k_4}
\chi_1 e^{i(k_2-k_3)} + \beta_{k_1}^\dag \beta_{k_2}^\dag \alpha_{k_3} \alpha_{k_4}
\chi_1 e^{i(k_1-k_4)} \right \}
\nonumber \\
& & \delta B_{p,p_y=0}^{\rm 6,leg} = \frac{1}{N^{5/2}}
\sum_{k_1 \ldots k_6} 2 \delta(p \!+\! k_1 \!+\! k_2 \!+\! k_3 \!-\! k_4
\!-\! k_5 \!-\! k_6) \nonumber \\
& & \qquad \left \{ \alpha_{k_1}^\dag \alpha_{k_2}^\dag \beta_{k_3}^\dag \beta_{k_4}
\beta_{k_5} \alpha_{k_6} e^{i(k_2-k_4-k_6)} \right.
\nonumber \\
& & \qquad\,\, \left. + \alpha_{k_1}^\dag \beta_{k_2}^\dag \beta_{k_3}^\dag \beta_{k_4}
\alpha_{k_5} \alpha_{k_6} e^{i(k_1+k_2-k_6)} \right \} \,.
\nonumber
\end{eqnarray}
for the in-phase component of the leg-polarization.
The out-of-phase component is very similar, only
the momentum $p$ has to be replaced with $(p+\pi)$, and 
in the term $\frac{1}{2} (e^{i(k_2-k_3)}+e^{i(k_1-k_4)})
\alpha_{k_1}^\dag \beta_{k_2}^\dag \beta_{k_3} \alpha_{k_4}$ 
of the 4--fermion component the parentheses have 
to be replaced with $(e^{i(k_2-k_3)}-e^{i(k_1-k_4)})$.

\begin{figure}[t]
\begin{center}\epsfig{figure=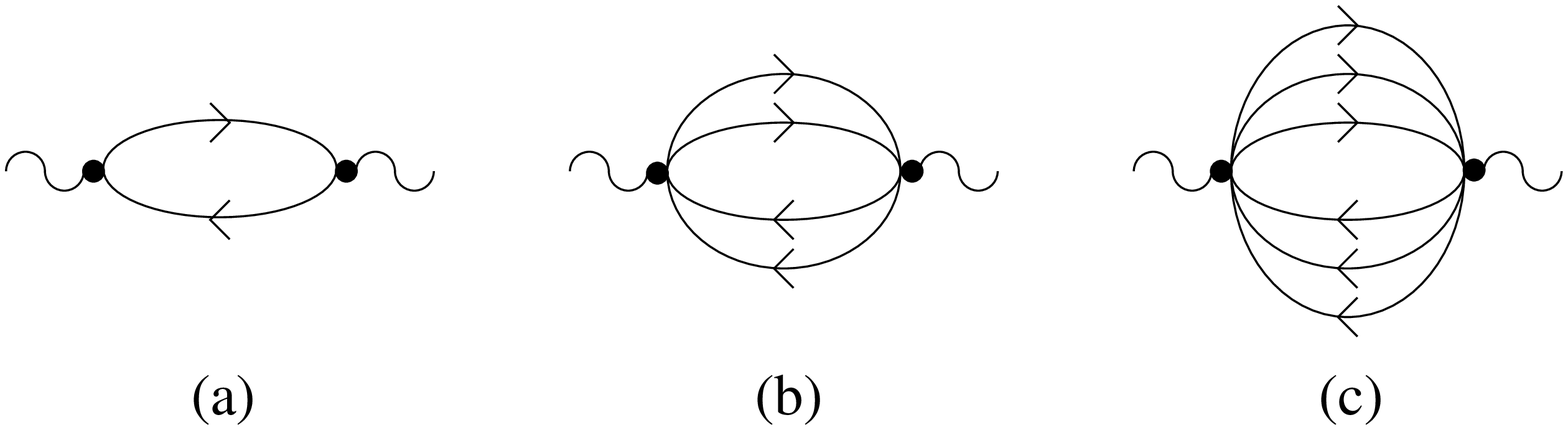,width=7cm,clip=}
\end{center}
\caption{
Mean-field weights for the spin--flip correlation.
They correspond to the evaluation of one, two and three
noninteracting particle-hole 
lines for the excitation of 2--, 4--
and 6--fermions displayed in (a), (b) and (c) respectively.}
\label{fig:WeightMF}  
\end{figure}

In order to estimate the contribution of processes involving the
excitation of a different number of fermions, we compare the weights 
\begin{equation}
\label{eq:weight}
W^{n}_{p_y}=-\int_0^\infty d\omega \frac{1}{N} \sum_{p}
\langle \langle f_{p} \delta B_{-p,p_y}^{n}; \delta B_{p,p_y}^{n} 
\rangle \rangle  \,,
\end{equation}
which are obtained by integrating the $n$=2, 4 and 6 fermion 
part to the leg-correlation functions over
frequency $\omega$ and momentum $p$ using form 
factors $f_{p}=1$ and $f_{p}=\sin^4(p/2)$.
To keep this evaluation as simple as possible the correlation
functions are evaluated in mean-field theory,
i.e., they are replaced by noninteracting particle-hole 
lines as indicated in Fig.~\ref{fig:WeightMF}. 
The frequency integrals in Eq.~(\ref{eq:weight}) can be eliminated
using:
\begin{eqnarray}
\pi \prod_j a_j &=&
- {\rm Im} \int_0^\infty \! d\omega \, e^{i \omega 0^+}
\prod_j \frac{1}{2 \pi i}
\int_{-\infty}^{\infty} \! d\Omega_j \times \\
& & 
\delta (\omega - \sum_j \Omega_j) (\frac{a_j}{\Omega_j-\epsilon_j+i0^+}
             +\frac{b_j}{\Omega_j+\epsilon_j-i0^+})
\nonumber
\end{eqnarray}
where $a_j/(\Omega_j \mp \epsilon_j \pm i0^+)$ are general expressions for
retarded and advanced Greens functions.
The remaining momentum integrals can be easily evaluated
numerically.

\begin{table}[b]
\caption{
Mean-field weights
of the 2--, 4-- and 6--fermion contributions to the in-phase ($p_y=0$)
and the out-of-phase ($p_y=\pi$) component assuming noninteracting particle-hole 
lines. For the 2--fermion contribution
the weight of the RPA-evaluation is given in parentheses.}
\label{table:Weights}
\medskip
\begin{tabular}{ccccc}
  & $p_y=0$ & $p_y=0$ & $p_y=\pi$ & $p_y=\pi$ \\
  & $f_p=1$ & $f_p\!=\sin^4(p/2)$ & $f_p=1$ & $f_p\!=\sin^4(p/2)$ \\
\hline
$n=2$  & 1.88  & 0.95  & 0.25  & 0.17  \\
       & (2.56) & (1.47) & (0.25) & (0.17) \\
$n=4$  & 0.59 & 0.17 & 0.45 & 0.17 \\
$n=6$  & 0.01 & 0.01 & 0.01 & 0.01 \\
\end{tabular}
\end{table}

The resulting weights of the 2--, 4-- and 6--fermion
processes are displayed in Table~\ref{table:Weights}.
The 4-- and 6--fermion processes make up only 24\% (15\%) of the
in-phase part of the spin-flip correlation function when a form factor
of $f_{p}=1$ ($f_{p}=\sin^4(p/2)$) is used. On the other hand, they generate
a major contribution to the out-of-phase component, i.e., 65\% (50\%).
Note, that these values refer only to noninteracting
particle-hole propagators and may be changed when the evaluation is
improved. For example the weight of the 2-fermion contribution to the
in-phase-component is enhanced 
when the particle-hole propagator is evaluated in RPA. In
Table~\ref{table:Weights} these RPA-weights are added in parentheses.

Since the 4-- and 6--fermion processes contribute only to the
continuum excitations, they will increase the high-energy
weight and therefore improve the consistency with the experimental
spectrum for polarization along the legs in
Fig.~\ref{fig:sigmaJWexp}. However, the consideration of 
4-- and 6--fermion processes will presumably not suffice to
account fully for the large high-energy weight observed experimentally.
One reason is, that the investigated spin-ladder compound
${\rm La_{5.2}Ca_{8.8}Cu_{24}O_{41}}$ corresponds not exactly to
an isotropic spin ladder $J=J_\perp$ but should rather be modeled by
$J\approx 1.3 J_\perp$ including a finite 4-spin cyclic exchange
interaction of $J_{cyc} \approx 0.2 J_\perp$.~\cite{condmat} The
presence of a finite cyclic spin exchange $J_{cyc}$ and the anisotropy
of the exchange couplings $J/J_\perp>1$ 
will increase the amount of high-energy continuum
excitations.~\cite{tobepublished} On the other 
hand it might also be necessary to include higher order vertex
corrections. They may shift some weight from the
bound state to the continuum excitations, because the RPA
overestimates the interaction strength as discussed in
Sec.~\ref{sec:SpinFlipCorrFuncLadder}. 
They may also help to restore the symmetries which are  broken artificially
by the meander path as discussed in Appendix~\ref{app:SymmetryBreaking}.

\section{Conclusions}
\label{sec:conclusions}

In this paper we have discussed an approach
to obtain dynamic correlation functions in low dimensional quantum
spin systems based on the Jordan-Wigner transformation. 
We have shown, that the application of standard perturbation
theory to the new fermionic operators is reasonable even in the
high-energy range. We have tested our approach by calculating the
dynamical spin-flip correlation function for the 1d spin chain, which
corresponds to the phonon assisted magnetic contribution to the optical
conductivity,  and we have found good agreement with the
experimental spectrum of ${\rm Sr_2CuO_3}$.

Using a meander-path like arrangement of the spin operators
we have extended this approach to the two-leg
$S=\frac{1}{2}$--ladders. In contrast to the 1d spin chain, however,
phase factors remain in the Hamiltonian.  
Expanding these phase factors creates new interaction terms,
which can be treated within standard perturbation theory.
We have calculated the one-particle dispersion in a mean-field
treatment based on nearest-neighbor bond amplitudes and 
the optical conductivity within an extended RPA-scheme.

For polarization along the rungs we find good agreement with the
experimental spectrum of the spin-ladder compound
${\rm La_{5.2}Ca_{8.8}Cu_{24}O_{41}}$.
Spin-flip processes on the rungs are represented quite reliably in our approach,
because two spins on the same rung are also neighboring along the meander path and
therefore no phase factor is present in the corresponding spin-flip operator.
For spin-flip processes along the legs, however, a phase factor
appears, which results in the excitation of 4-- and 6--fermions
in addition to the 2--fermion processes. Evaluation of the 2--fermion
contribution to the spin-flip correlation function for polarization
along the legs yields good results with respect to the
$S$=0--bound state. The weight of the high-energy continuum, however, is 
underestimated. To some extent this can be compensated by the
consideration of 4-- and 6--fermion excitations.

In principle our approach can be generalized straightforwardly to spin-ladders
with more than two legs. However one should think about 
a less refined treatment of the phase factors in order to reduce the
increasing number of interaction terms resulting from the longer range of the
phase factors. 

\section*{Acknowledgments}
We would like to thank  M.~Gr\"uninger and Q.~Yuan for stimulating discussions.
This project was supported by the DFG (SFB 484)
and by the BMBF (13N6918/1).

\appendix
\section{Strong coupling limit}
\label{app:DimLimit}

In this appendix we 
demonstrate that our mean-field treatment for
the Jordan-Wigner fermions, which is based on the meander path, can
reproduce the correct strong coupling limit $\epsilon_k=J_\perp+J\cos k$.
This is not true for the zigzag path within a straightforward extension
of our mean-field approach. 
\subsection{Meander-path}
Expanding around the rung-dimer limit we obtain:
\begin{itemize}
\item zeroth order: $(J/J_\perp)^0$ \\
For $J=0$ the off-diagonal part of the
mean-field Hamiltonian~(\ref{eq:HMF}) reduces to
\begin{equation}
\gamma_k=J_\perp (\frac{1}{2}-\chi_0) \, .
\end{equation}
The resulting Hamiltonian can be
diagonalized easily using $u_k=v_k=1$ in Eq.~(\ref{eq:diagtrafo})
which yields for the nearest neighbor bond amplitudes
\begin{eqnarray}
\chi_0\!&=&\!-\frac{1}{2N} \sum_k u_k^2 \!=\! - \frac{1}{2} \,, \,\,
\chi_1\!=\!-\frac{1}{2N} \sum_k u_k^2 e^{-ik}\!=\!0 \, ,
\nonumber \\
\chi_2\!&=&\!-\frac{1}{2N} \sum_k v_k^2 e^{-ik} \!=\! 0 \, .
\end{eqnarray}
This is the limit of rung dimers and the correct value for the energy of 
a single rung-triplet excitation is obtained as
\begin{equation}
\epsilon_k = \vert \gamma_k \vert = J_\perp \,.
\end{equation}
\item first order: $(J/J_\perp)^1$ \\
Re-substituting the zeroth order bond amplitudes in Eq.~(\ref{eq:gamma}) yields
$\gamma_k=J_\perp + J (\cos k + i \sin k)$. Expanding the resulting dispersion
to first order in $J/J_\perp$ one obtains the correct strong
coupling limit:
\begin{eqnarray}
\label{eq:limitJ0}
\epsilon_k = \vert \gamma_k \vert 
           &=& J_\perp \sqrt{1 + 2 \frac{J}{J_\perp} \cos k +
                       (\frac{J}{J_\perp})^2}
\nonumber \\
           &\approx& J_\perp+J\cos k \, .
\end{eqnarray}
This expression contains already all terms to order $J/J_\perp$.
This can be easily seen when the above results are inserted into
Eq.~\ref{eq:bondamp} for the bond amplitudes. To order
$J/J_\perp$ one obtains for the diagonalization transformation
$u_k=\gamma_k/\vert \gamma_k \vert \approx 1 + i J/J_\perp \sin k$
and consequently for the bond amplitudes 
$\chi_0=-1/2$ and $\chi_1=-\chi_2=J/(4J_\perp)$.
Whereas $\chi_0$ remains unchanged, $\chi_1$ and $\chi_2$
are proportional to $J/J_\perp$ and therefore 
contribute to $\gamma_k$ and $\epsilon_k$ only in second
order. 
\end{itemize}

\subsection{Zigzag--path}
In this subsection we discuss the mean field treatment for the
zigzag path. In particular we show that it is does not reproduce the
correct strong coupling limit.

For an illustration of the zigzag path see
Fig.~\ref{fig:JordanWignerPfade}(b). Here, the underlying sublattice
structure is very simple, each leg corresponds to a
sublattice. Choosing, e.g., in the Jordan-Wigner transformation,
Eq.~(\ref{eq:JWtrafo}), $S_{i,\alpha}$ on
the lower leg and $S_{i,\beta}$ on the upper leg of
Fig.~\ref{fig:JordanWignerPfade}(b), the XY-part of Hamiltonian along
the legs becomes 
\begin{eqnarray}
H_{\rm XY,leg} &=& J \sum_i \left \{ \frac{1}{2} \left [
\alpha_i^\dag \alpha_{i+1} (1 - 2\beta_i^\dag \beta_i) \right. \right.
\\
& & \qquad\quad\,\, + \left. \left.
    \beta_i^\dag \beta_{i+1} (1 - 2\alpha_{i+1}^\dag \alpha_{i+1})
+ {\rm H.c.}  \right ] \right \} \,. \nonumber
\end{eqnarray}
Here the expansion of the phase factor yields only 4--operator
terms. Taking into account all nearest and next nearest
neighbor bond amplitudes:
\begin{eqnarray}
\chi_0 &=& \langle \beta_i^\dag \alpha_i \rangle \, , \,\,
\chi_1=\langle \alpha_i^\dag \alpha_{i+1} \rangle 
\\
\chi_2 &=& \langle \beta_i^\dag \beta_{i+1} \rangle \, , \,\,
\chi_3=\langle \beta_i^\dag \alpha_{i+1} \rangle \, , \,\,
\chi_4=\langle \alpha_i^\dag \beta_{i+1} \rangle 
\nonumber
\end{eqnarray}
one obtains the following mean-field Hamiltonian
\begin{equation}
\label{eq:HMFzigzag}
H_{\rm MF} = \sum_k \left \{ \gamma_{0,k} \alpha_k^\dag \beta_k 
             - \gamma_{\alpha,k} \alpha_k^\dag \alpha_k
	     - \gamma_{\beta,k} \beta_k^\dag \beta_k + H.c. \right \}
\end{equation}
with
\begin{eqnarray}
\gamma_{0,k} &=& J_\perp (\frac{1}{2}-\chi_0)
                    + 2J(\chi_0 e^{ik} + \chi_3)
\nonumber \\
\gamma_{\alpha,k} &=& J \chi_1 e^{ik} \,\, , \,\, 
\gamma_{\beta,k} = J \chi_2 e^{ik} \, .
\end{eqnarray}
For $J=0$ this equals the mean-field Hamiltonian for
the meander path, Eq.~(\ref{eq:HMF}), and therefore the same values for the bond amplitudes 
$\chi_0=-\frac{1}{2} \, , \, \chi_1=\chi_2=\chi_3=0$ are obtained.
Reinserting these values into Eq.~(\ref{eq:HMFzigzag}) yields
\begin{equation}
\gamma_{0,k}=J_\perp-J(\cos k + i \sin k) \, , \,
 \gamma_{\alpha,k}=\gamma_{\beta,k}=0 \, .
\end{equation}
The corresponding dispersion is to first order in $J/J_\perp$
\begin{eqnarray}
\epsilon_k=\vert \gamma_{0,k} \vert &=&
          J_\perp \sqrt{1 - 2 \frac{J}{J_\perp} \cos k +
                       (\frac{J}{J_\perp})^2}
\nonumber \\
&\approx& J_\perp - J \cos k
\end{eqnarray}
which obviously differs from the correct strong coupling
limit~(\ref{eq:limitJ0}).

\section{RPA equations}
\label{app:RPAequations}
The RPA-equations for the particle-hole propagators of the
spinless Jordan-Wigner fermions can be obtained by considering all
possible vertex configurations of the RPA-Hamiltonian~(\ref{eq:HRPA}).
The explicit form for the renormalized particle-hole propagators is:
\begin{eqnarray}
B^{\mu,\nu}_{\gamma^\dag \! \rho, \delta \sigma^\dag} \!\!
= \tilde b^{\mu,\nu}_{\gamma^\dag \! \rho, \delta \sigma^\dag} \!\!
&+& \!\! \sum_{\lambda=0}^{2} \! \left \{ 
b^{\mu,3 \lambda}_{\gamma^\dag\! \alpha, \delta \alpha^\dag}
B^{\lambda,\nu}_{\beta^\dag \! \rho, \beta \sigma^\dag}
\!+ b^{\mu,1 \lambda}_{\gamma^\dag\! \alpha, \delta \beta^\dag}
B^{\lambda,\nu}_{\alpha^\dag\! \rho, \beta \sigma^\dag}
\right. \nonumber \\
& & \quad \!+ \, b^{\mu,6 \lambda}_{\gamma^\dag\! \alpha, \delta \beta^\dag}
B^{\lambda,\nu}_{\beta^\dag\! \rho, \alpha \sigma^\dag}
\!+ b^{\mu,4 \lambda}_{\gamma^\dag\! \beta, \delta \alpha^\dag}
B^{\lambda,\nu}_{\beta^\dag\! \rho, \alpha \sigma^\dag}
\nonumber \\
& & \quad + \left.\! b^{\mu,5 \lambda}_{\gamma^\dag\! \beta, \delta \alpha^\dag}
B^{\lambda,\nu}_{\alpha^\dag\! \rho, \beta \sigma^\dag}
\!+ b^{\mu,2 \lambda}_{\gamma^\dag\! \beta, \delta \beta^\dag}
B^{\lambda,\nu}_{\alpha^\dag\! \rho, \alpha \sigma^\dag}
\right \}
\nonumber \\
\end{eqnarray}
For simplicity the momentum and frequency indices $p$ and $\omega$
have been omitted. The noninteracting particle-hole propagators $\tilde b$ and $b$,
including the appropriate form factors from the internal interaction vertices,
 are defined as:
\begin{eqnarray}
\tilde b^{\mu,\nu}_{\gamma^\dag \! \rho, \delta \sigma^\dag} &=&
\frac{1}{N} \sum_k f_k^\mu f_k^\nu \,
b^0_{\gamma^\dag \! \rho, \delta \sigma^\dag} (p,k,\omega)
\nonumber \\
b^{\mu,s \lambda}_{\gamma^\dag \! \rho, \delta \sigma^\dag} &=&
\frac{1}{N} \sum_k f_k^\mu g^{(s,\lambda)}(p,k)  \,
b^0_{\gamma^\dag \! \rho, \delta \sigma^\dag} (p,k,\omega)
\end{eqnarray}
with form factors
\begin{equation}
f_k^0=1 \,\,\,\, , \,\,\,\, f_k^1=e^{ik} \,\,\,\, , \,\,\,\,
f_k^2=e^{-ik} \,,
\end{equation}
and
\begin{eqnarray}
g^{(1,0)}(p,k) &=& - J_\perp f_k^0 + 2 J \chi_0 (1 +e^{-ip}) f_k^2
\nonumber \\
g^{(1,1)}(p,k) &=& J \left [ 2 \chi_0 (1+e^{ip}) f_k^0
                -2 \chi_1 e^{ip} f_k^1 - (1+4 \chi_2) f_k^2 \right ]               
\nonumber \\
g^{(1,2)}(p,k) &=& -J \left ( f_k^1 + 2\chi_1 e^{-ip} f_k^2 \right )
\nonumber \\
g^{(2,0)}(p,k) &=& \left [ J_\perp + J \left ( e^{-ip} + (1+4 \chi_2) 
e^{ip} \right ) \right ] f_k^0
\nonumber \\
& & \qquad - \, 2 J \chi_0 e^{ip} f_k^1 - 2 J \chi_0 f_k^2
\nonumber \\
g^{(2,1)}(p,k) &=& 2 J \left ( - \chi_0 e^{ip} f_k^0 + \chi_1 e^{ip}
f_k^1 \right )
\nonumber \\
g^{(2,2)}(p,k) &=& 2 J \left ( - \chi_0 f_k^0 + \chi_1 e^{-ip}
f_k^2 \right )
\nonumber \\
g^{(3,0)}(p,k) &=& \left [ J_\perp + J \left ( e^{ip} + (1+4 \chi_2)
e^{-ip} \right ) \right ] f_k^0
\nonumber \\
& & \qquad - \, 2 J \chi_0 f_k^1 - 2 J \chi_0 e^{-ip} f_k^2
\nonumber \\
g^{(3,1)}(p,k) &=& 2 J \left ( - \chi_0 f_k^0 + \chi_1 e^{ip}
f_k^1 \right )
\nonumber \\
g^{(3,2)}(p,k) &=& 2 J \left ( - \chi_0 e^{-ip} f_k^0 + \chi_1 e^{-ip}
f_k^2 \right )
\nonumber \\
g^{(4,0)}(p,k) &=& - J_\perp f_k^0 + 2 J \chi_0 (1 +e^{ip}) f_k^1
\nonumber \\
g^{(4,1)}(p,k) &=& -J \left ( 2 \chi_1 e^{ip} f_k^1 + f_k^2 \right )           
\nonumber \\
g^{(4,2)}(p,k) &=& J \left [ 2 \chi_0 (1+e^{-ip}) f_k^0
                - (1+4 \chi_2) f_k^1 \right.
\nonumber \\
& & \qquad\qquad\qquad\quad\quad\,\, \left. - \, 2 \chi_1 e^{-ip} f_k^2 \right ] 
\nonumber \\
g^{(5,0)}(p,k) &=& g^{(6,0)}(p,k) = 2 J \chi_1 (e^{ip}+e^{-ip}) f_k^0
\nonumber \\
g^{(5,1)}(p,k) &=& g^{(6,1)}(p,k) = - 2 J \chi_1 f_k^2
\nonumber \\
g^{(5,2)}(p,k) &=& g^{(6,2)}(p,k) = - 2 J \chi_1 f_k^1
\end{eqnarray}
The noninteracting 
particle-hole propagators $b^0$ are given as:
\begin{eqnarray}
b^0_{\alpha^\dag \! \alpha, \alpha \alpha^\dag}
&=& b^0_{\beta^\dag \! \beta, \beta \beta^\dag}
= b^0_{\alpha^\dag \! \alpha, \beta \beta^\dag}
= b^0_{\beta^\dag \! \beta, \alpha \alpha^\dag}=
\nonumber \\
&=& \frac{1}{4} \left( g_1^0(p,k,\omega) - g_2^0(p,k,\omega) \right )
\nonumber \\
b^0_{\beta^\dag \! \alpha, \alpha \alpha^\dag} 
&=& b^0_{\beta^\dag \! \alpha, \beta \beta^\dag}
= \frac{1}{4} u_{p+k}^2 \left( g_1^0(p,k,\omega) + g_2^0(p,k,\omega) \right )
\nonumber \\
b^0_{\alpha^\dag \! \beta, \alpha \alpha^\dag} 
&=& b^0_{\alpha^\dag \! \beta, \beta \beta^\dag}
= \frac{1}{4} v_{p+k}^2 \left( g_1^0(p,k,\omega) + g_2^0(p,k,\omega) \right )
\nonumber \\
b^0_{\alpha^\dag \! \alpha, \alpha \beta^\dag} 
&=& b^0_{\beta^\dag \! \beta, \alpha \beta^\dag}
= - \frac{1}{4} u_k^2 \left( g_1^0(p,k,\omega) + g_2^0(p,k,\omega) \right )
\nonumber \\
b^0_{\alpha^\dag \! \alpha, \beta \alpha^\dag} 
&=& b^0_{\beta^\dag \! \beta, \beta \alpha^\dag}
= - \frac{1}{4} v_k^2 \left( g_1^0(p,k,\omega) + g_2^0(p,k,\omega) \right )
\nonumber \\
b^0_{\alpha^\dag \! \beta, \alpha \beta^\dag}
&=& - \frac{1}{4} u_k^2 v_{p+k}^2 \left( g_1^0(p,k,\omega) - g_2^0(p,k,\omega) \right )
\nonumber \\
b^0_{\beta^\dag \! \alpha, \beta \alpha^\dag}
&=& - \frac{1}{4} v_k^2 u_{p+k}^2 \left( g_1^0(p,k,\omega) - g_2^0(p,k,\omega) \right )
\nonumber \\
b^0_{\alpha^\dag \! \beta, \beta \alpha^\dag}
&=& - \frac{1}{4} v_k^2 v_{p+k}^2 \left( g_1^0(p,k,\omega) - g_2^0(p,k,\omega) \right )
\nonumber \\
b^0_{\beta^\dag \! \alpha, \alpha \beta^\dag}
&=& - \frac{1}{4} u_k^2 u_{p+k}^2 \left( g_1^0(p,k,\omega) - g_2^0(p,k,\omega) \right )
\end{eqnarray}
with
\begin{eqnarray}
g_1^0(p,k,\omega) &=& \frac{1}{\omega-\epsilon_k-\epsilon_{p+k}+i\delta}
\nonumber \\
g_2^0(p,k,\omega) &=&
\frac{1}{\omega+\epsilon_k+\epsilon_{p+k}-i\delta} \, .
\end{eqnarray}
Here, $\chi_0$, $\chi_1$ and $\chi_2$ are the bond amplitudes~(\ref{eq:bondamp}),
$u_k$, $v_k$ are the coefficients for the diagonalization~(\ref{eq:diagtrafo}) of the
mean-field Hamiltonian~(\ref{eq:HMF}) and $\epsilon_k$ is the
mean-field dispersion~(\ref{eq:Hdiag}).

\section{Role of the phase factor}
\label{app:PhaseFactor}

\begin{figure}[b!]
\begin{center}
\epsfig{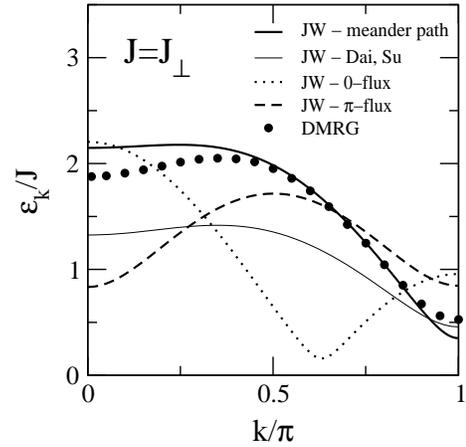}
\end{center}
\caption{Dispersion obtained for a flux-phase
approximation of the phase factor with 0-flux (dotted line) and
$\pi$-flux (dashed line) through a plaquette.
Thick solid line: mean-field dispersion obtained for the meander path in
Sec.~\protect{\ref{sec:MeanFieldLadder}} where the
correlations related to the phase factor have been taken into account.
Thin solid line: dispersion analogous to Dai and Su,\cite{DaiSu}
where the phase factor has been replaced by its average value.
Symbols: one-triplet dispersion obtained by DMRG for a $N$=80-site
ladder.\cite{condmat}}
\label{fig:DispersionOhnePhase}  
\end{figure}

In this appendix we briefly comment on the role of the ``phase factor''
in the mean-field evaluation of the triplet dispersion. 
The phase factor $e^{i\pi(n_{\beta_i}+n_{\alpha_{i+1}})}$ in
Hamiltonian~(\ref{eq:HeisenbergLadderJW}) 
was generated by products of spin operators with site
labels not in sequence along the one-dimensional meander path.
As the ``matrix elements'' of the phase operator are $\pm1$ 
one might speculate to find a reasonable mean field result
by replacing the operator uniformly by $\pm1$.
This corresponds to a flux phase treatment of the phase factor,
where the flux through a plaquette is chosen to be 0 and $\pi$,
respectively. Applying the same kind of mean field treatment
as in Sec.~\ref{sec:MeanFieldLadder} one obtains $\epsilon_k = |\gamma_k|$
for the mean field dispersion, where
\begin{displaymath}
\gamma_k=J_\perp (\frac{1}{2}\!-\!\chi_0)+J \cos k (1\!-\!\chi_1\!-\!\chi_2)
                                     +i J \sin k (\chi_2\!-\!\chi_1)
\end{displaymath}
corresponds to zero flux and
\begin{displaymath}
\gamma_k=J_\perp (\frac{1}{2}\!-\!\chi_0)-J \cos k (\chi_1\!+\!\chi_2)
                                     +i J \sin k (1-\chi_1\!+\!\chi_2)
\end{displaymath}
to a $\pi$-flux phase. The resulting dispersions are displayed in
Fig.~\ref{fig:DispersionOhnePhase} together
with the mean field dispersion of the meander-path,
where (thick solid line) the correlations related to
the phase factor are taken into account and where (thin solid line)
the phase factor has been replaced by its average value. For
comparison also the DMRG results for a $N$=80-site ladder~\cite{condmat}
are displayed. Obviously, the form of the dispersion 
improves considerably when the phase factor is treated in a more
adequate way. The simple replacement by a flux phase shows only poor
agreement with the DMRG result.
It is interesting to note, however, 
that the spin gap remains finite for all $J_\perp/J>0$
within the $\pi$-flux treatment of the phase factor, as has been
observed by Azzouz {\it et al.}~\cite{Azzouz}
The mean field evaluation of the phase factor (thin
solid line) improves 
the form of the dispersion at least qualitatively.
For a reasonable quantitative agreement with the exact dispersion,
however, it seems to be necessary to consider also the correlations
related to the phase factor (thick solid line).

\section{Symmetry breaking}
\label{app:SymmetryBreaking}

A problem of applying the Jordan-Wigner transformation by placing a path
throughout the ladder is that both the zigzag and the meander path
break symmetries of the original ladder configuration. This
problem has already been mentioned in the discussion of the
out-of-phase spin-flip correlation function in 
Sec.~\ref{sec:SpinFlipCorrFuncLadder}.
The sublattice structure underlying the meander path, see Fig.~\ref{fig:Sublattice},
restricts the translational symmetry along the legs to
translations of an even number of sites. Furthermore, it breaks the reflection
symmetry with respect to the center line crossing all rungs perpendicularly.
In MFA this gives rise to different values for the bond amplitudes 
$\chi_1=\langle \beta_i^\dagger \alpha_{i+1} \rangle=-0.2679$
and $\chi_2=\langle \alpha_i^\dagger \beta_{i+1} \rangle=0.1777$. Even
though the fermionic bond amplitudes have no direct physical
meaning, these discrepancies translate
into different expectation values for the products of neighboring spin
operators. When the spin operators are transformed to fermionic operators 
according to Eq.~(\ref{eq:JWtrafo}) and all contractions are
replaced by their mean-field values as before, we obtain for the
product of the $z$-components 
$\langle S_{i,\alpha}^z S_{i+1,\beta}^z \rangle = -|\chi_2|^2$,
$\langle S_{i,\beta}^z S_{i+1,\alpha}^z \rangle = -|\chi_1|^2$
and for the $xy$-components
$\frac{1}{2} \langle S_{i,\beta}^+ S_{i+1,\alpha}^-
+ S_{i,\beta}^- S_{i+1,\alpha}^+ \rangle = \chi_1$ and 
$\frac{1}{2} \langle S_{i,\alpha}^+ S_{i+1,\beta}^-
+ S_{i,\alpha}^- S_{i+1,\beta}^+ \rangle 
= 4 \chi_1 ( \chi_0^2 - \chi_1 \chi_2)$.
Therefore two symmetries are broken with respect to the spin
operators: (i) translational symmetry by an odd number of sites as
discussed before $\langle S_{i,\alpha} S_{i+1,\beta} \rangle \ne
\langle S_{i+1,\beta} S_{i+2,\alpha} \rangle$ and (ii) rotational
symmetry in spin space $\langle S_{i,\alpha}^z S_{i+1,\beta}^z \rangle
\ne \frac{1}{4} \langle S_{i,\alpha}^+ S_{i+1,\beta}^- +
S_{i,\alpha}^- S_{i+1,\beta}^+ \rangle$.

The physical interpretation of  breaking  the translational
symmetry is quite obvious. $\langle S_{i,\alpha} S_{i+1,\beta} \rangle \ne
\langle S_{i+1,\beta} S_{i+2,\alpha} \rangle$ signifies that a static
dimerization pattern is induced along the meander path. Its
amplitude, however, is very small $\langle S_{i+1,\beta} S_{i+2,\alpha} -
S_{i,\alpha} S_{i+1,\beta} \rangle = 0.037$. A staggered dimerization
pattern of this kind can, e.g., be induced by consideration of a 4-spin
cyclic exchange interaction~\cite{LaeuchliTroyer}
of about $J_{cyc}/J \approx 1-2$. Indeed, it has been found that the
inclusion of a cyclic spin exchange is necessary in order to obtain a
consistent description of the experimental data for
cuprate spin ladder compounds like
(La,Ca)$_{14}$Cu$_{24}$O$_{41}$.~\cite{condmat} The realistic value of
$J_{cyc}\approx0.2-0.27 J_\perp$, however, is not strong enough to induce a
static dimerization pattern. Nevertheless the correlation length for
dimerization fluctuations increases with cyclic spin exchange and 
the dip in the one--triplet 
dispersion is reduced.~\cite{condmat} 
This could to some extent explain
the good agreement we find for our Jordan-Wigner calculation 
with the experimental spectra of (La,Ca)$_{14}$Cu$_{24}$O$_{41}$, as
our artificial symmetry breaking seems to 
imitate some of the effects of a cyclic spin exchange.


\end{document}